# What is Shear Wave

Peng SHI[a, b]

[a] Department of Astronautics and Mechanics, Harbin Institute of Technology, Postbox 344, 92 West Dazhi Street, Harbin, 150001, P. R. China

[b] Department of engineering mechanics, School of civil engineering and architecture, Xi'an University of Technology, No. 5, Jinhua South Road, Xi'an, 710048, P. R. China

Corresponding author: Peng SHI sp198911@outlook.com;

***Abstract.*** This study shows that the traditional definition of shear wave breaks the shear stress reciprocity. By analyzing the displacement field for shear wave and the motion equation of material element, it is found that the displacement field is related to local rigid body rotation, and the local rigid body rotation cannot be balanced by the stress state assumed in classical continuum mechanics. It is also found that the displacement field caused by shear deformation, which is neither divergence nor rotation, is replaced with that caused by local rigid body rotation during the derivation of wave equation. This indicates that the definition of shear wave is beyond the basic assumption of continuum mechanics. The study modified the constitutive relation and elastic tensor based on that the traditionally defined shear wave is objective. The motion equation corresponding to shear wave is also derived. It is concluded from the definition of traditional shear wave that the local rigid body rotation should contribute stress and the shear stress reciprocity is not prerequisite for nonpolar continuum.

**Introduction**

In order to study the forces and movements of materials occupying a certain space, the macroscopic medium is considered to satisfy the continuum hypothesis, in which the structures of real fluid and solid are considered to be perfectly continuous and are paid no attention to their molecular structure [1,2]. Then, continuum mechanics is established to describe the forces and movements of continuum materials based on the basic hypothesis. The continuum mechanics considers that the dynamics of continuums still satisfy Newton's law, and the element translation only needs to be considered when describing its conservation of moment of momentum [3-5]. This means that the possible rotation of nonzero-volume elements constituting continuum is ignored, and the element is treated as particle that only has translation. Based on the above hypothesis, the Cauchy stress tensor, also known as the true stress tensor, is considered to fully describe the stress state at a point under the current configuration and to be a second-order symmetric stress tensor for non-polar materials [6-9].

Continuum mechanics is widely used to describe macroscopic mechanical properties of macroscopic media, such as elastomers [7]. For a linear elastomer, the strain tensor is considered to be a second-order symmetric tensor due to shear stress reciprocity, which can be expressed by the symmetrical component of the displacement gradient [1, 10]. Bringing the stress-strain relationship and geometric equation of elastomers into the equation of momentum conservation in the differential form, the wave equation (or the displacement equation of motion) can be obtained [6,7,9]. The propagation of elastic waves in elastomers is considered to be governed by the equation. However, the displacement field in the traditionally defined shear wave is related to local rigid body rotation, which means that the rotation of material element is considered. This is conflicted with the assumption that the material element forming elastomers can be treated as a particle whose rotation can be neglected in classical continuum mechanics. Hence, the theory of elasticity based on classical continuum mechanics shouldn't have obtained the traditionally defined shear wave. If the definition of traditional shear wave is objective, the shear stress reciprocity should be broken to balance the rotation of material element in the traditionally defined shear wave, and the local rigid body rotation, which is the antisymmetric part of the displacement gradient, should be considered to contribute the antisymmetric part of stress tensor. This indicates that there are mistakes in the derivation of wave equation. The point doesn't seem to have been realized so far, which motivates me to do this study.

In this study, I will point out what mistakes have been made during the derivation of wave equation under shear stress reciprocity. I will also show that the motion of material element described by the traditionally considered wave equation is different from that described by the traditionally defined shear wave. In order to derive the shear wave equation under the assumption of classical continuum mechanics and give the constitutive relation and elastic tensor based on that the traditionally defined shear wave the displacement equation of motion from the traditional definition of shear wave, the isotropic elastomer is used as an example.

**Derivation of the displacement equation of motion**

Only considering the element translation, for elastomer with small deformation, the equation of motion in differential form is expressed as [7]

$$\rho \frac{\partial^2 \boldsymbol{u}}{\partial t^2} - \nabla \cdot \boldsymbol{\sigma} - \boldsymbol{f} = 0, \tag{1}$$

here, $\partial/\partial t$ is the time derivative, $\nabla$ is the vector operator del, $\boldsymbol{\sigma}$ is a symmetric second-order stress tensor, $\boldsymbol{f}$ is body force which is an irrotational field, $\rho$ is the mass density, $\boldsymbol{u}$ is the displacement. Equation (1) is obtained by Gauss's theorem. Therefore, an element particle on the microscopic level is a material volume that the dynamic state of continuum in its subregion is constant [11].

Since classical continuum mechanics doesn't consider the rotation of element, the shear stress is reciprocal. Therefore, the strain tensor in the theory of elasticity, which describes the deformation of elastomer under force, must also satisfy shear strain reciprocity. In the theory of elasticity, the symmetric part of the displacement gradient is used to express the strain, and the anti-symmetric part of the displacement gradient is treated as the local rigid body rotation without contributing stress [7,8]. Hence, the stress-strain relationship of elastomer is expressed as follows:

$$\boldsymbol{\sigma} = \boldsymbol{C} : \boldsymbol{e}, \tag{2}$$

here, $\boldsymbol{e}$ is a second-order symmetric strain tensor, and $\boldsymbol{C}$ is a fourth-order elastic tensor. For isotropic elastomers, the stress-strain relationship in component form is expressed as [7,9]

$$\sigma_{ij} = \lambda e_{kk} \delta_{ij} + 2\mu e_{ij}, \tag{3}$$

where, $\lambda$ and $\mu$ are Lamé constants, $\delta$ is the Kronecker delta symbol. The strain tensor is expressed with displacement as:

$$\boldsymbol{e} = \frac{1}{2}\left(\nabla \boldsymbol{u} + \nabla \boldsymbol{u}^T\right), \tag{4}$$

Here, $\boldsymbol{u}^T$ is the transposition of displacement. The relative displacement between two points in the neighborhood can be expressed as [12]:

$$\boldsymbol{u}(\boldsymbol{x} + \delta \boldsymbol{r}) = \boldsymbol{u}(\boldsymbol{r}) + \boldsymbol{e} \cdot \delta \boldsymbol{r} + \boldsymbol{\Omega} \cdot \delta \boldsymbol{r}, \tag{5}$$

$$\boldsymbol{\Omega} = \frac{1}{2}\left(\nabla \boldsymbol{u} - \nabla \boldsymbol{u}^T\right), \tag{6}$$

Where, $\delta\boldsymbol{r}$ is displacement vector of two points in the neighborhood, $\boldsymbol{\Omega}$ is rotation tensor, which is an anti-symmetric tensor. It can be seen form Equation (5) that the displacement field $\boldsymbol{u}^S$ should be an irrotational field if the displacement is only caused by deformation:

$$\nabla \times \boldsymbol{u}^S = \boldsymbol{\varepsilon} : \boldsymbol{\Omega} = 0, \tag{7}$$

$$\boldsymbol{u}^S(\boldsymbol{x} + \delta\boldsymbol{r}) = \boldsymbol{u}^S(\boldsymbol{r}) + \boldsymbol{e} \cdot \delta\boldsymbol{r}, \tag{8}$$

where, $\varepsilon$ is the permutation symbol. If it is further assumed that the displacement field is caused by pure shear deformation, the displacement field should be neither rotational nor divergent:

$$\nabla \cdot \boldsymbol{u}^S = tr(\nabla \boldsymbol{u}^S) = tr(\boldsymbol{\varepsilon}) = 0. \tag{9}$$

This is different from the traditional understanding of displacement field, which is considered to be either divergence field or curl field [7,9,12,13].

Substituting Equations (2)-(4) into Equation (1) and ignoring the body force, the equation of motion expressed by displacement can be obtained:

$$(\lambda + \mu)\nabla(\nabla \cdot \boldsymbol{u}) + \mu\nabla^2 \boldsymbol{u} = \rho \frac{\partial^2 \boldsymbol{u}}{\partial t^2}. \tag{10}$$

This equation is also called elastic wave equation or Navier's equation.

**Derivation of shear wave equation**

The stress-strain relationship (Equation (2)) is often called generalized Hooke's law. Then, Equation (10) is also regarded as the motion equation of the generalized spring proton model. As the local rigid body rotation doesn't contribute force, Equation (10) will not be able to balance the motion of the element if it rotates. This means that the displacement in Equation (10) should be only related to the deformation. Therefore, the displacement field in Equation (10) should be irrotational. When only shear deformation occurs to the elastomer, Equation (10) can be reduced to the following equation:

$$\mu \nabla^2 \boldsymbol{u} = \rho \frac{\partial^2 \boldsymbol{u}}{\partial t^2}. \tag{11}$$

In this case, the displacement field in Equation (11) should be neither rotational nor divergent. The displacement field is different from the traditional understanding of displacement field caused by shear deformation which is a curl field. From Equation (11), it is obtained that a shear wave should be defined as a wave propagating at shear wave velocity, in which the displacement field caused by shear deformation is neither rotational nor divergent. In the case of a two-dimensional plane wave ($\partial/\partial x_3 = 0$ and $u_3 = 0$), the shear wave equations should be expressed as:

$$\mu \left( \frac{\partial^2 u_1}{\partial x_1^2} + \frac{\partial^2 u_1}{\partial x_2^2} \right) = \rho \frac{\partial^2 u_1}{\partial t^2}, \tag{12}$$

$$\mu \left( \frac{\partial^2 u_2}{\partial x_1^2} + \frac{\partial^2 u_2}{\partial x_2^2} \right) = \rho \frac{\partial^2 u_2}{\partial t^2}, \tag{13}$$

here, $u_1$ and $u_2$ are the displacement components along the direction of $x_1$ and $x_2$, respectively. To some extent, the shear wave can also be regarded as a special longitudinal wave with shear wave velocity and vertical strain not equal to zero. At this time, the vertical strain is constant along the vertical direction of the wave propagation. From the definition of shear wave in Equation (11), the displacement perpendicular to the propagation direction is non-zero. However, the displacement field need to be neither rotational nor divergent. Further assuming that the wave propagates along the $x_1$-direction, the shear wave equations

should be expressed as:

$$\mu \frac{\partial^2 u_1}{\partial x_1^2} = \rho \frac{\partial^2 u_1}{\partial t^2}, \quad (14)$$

$$\frac{\partial u_1}{\partial x_1} = -\frac{\partial u_2}{\partial x_2}. \quad (15)$$

If a shear wave with reciprocal shear stress exists, it should be as expressed with Equations (14) and (15) and should be different from the traditional definition of shear wave.

In the traditional understanding of displacement field, the displacement field is assumed as a superposition of curl field and divergence field, and the following formula is considered to be always true [7,9,12,13]:

$$\nabla^2 \boldsymbol{u} = \nabla(\nabla \cdot \boldsymbol{u}) - \nabla \times \nabla \times \boldsymbol{u}. \quad (16)$$

Substituting Equation (16) in to Equation (10), the following formula is obtained:

$$(\lambda + 2\mu)\nabla(\nabla \cdot \boldsymbol{u}) - \mu \nabla \times \nabla \times \boldsymbol{u} = \rho \frac{\partial^2 \boldsymbol{u}}{\partial t^2}. \quad (17)$$

Equation (17) is considered to be an alternative form of the displacement equation of motion. As mentioned above, the local rigid body rotation does not contribute to force. The displacement in Equation (10) should be independent to the rotation of local rigid body. Therefore, Equation (16) shouldn't include the rotational part. on the other hand, it is seen from Equation (16) that the displacement field caused by shear deformation cannot be shown in the wave equation. With Equation (16), the shear deformation, which produces a displacement field without rotation or divergence, is actually replaced by the local rigid body rotation. Hence, the wave propagating at shear wave velocity defined in Equation (10) is different from the shear wave expressed with Equation (17). Equation (17) can be also written as:

$$(\lambda + 2\mu)\nabla(\nabla \cdot \boldsymbol{u}) - \nabla \times \left(\frac{1}{2}\boldsymbol{\varepsilon} : 2\mu\boldsymbol{\Omega}\right) = \rho \frac{\partial^2 \boldsymbol{u}}{\partial t^2}. \qquad (18)$$

From Equation (18), it is seen that the local rigid body rotation contributes to stress. When the displacement in Equation (18) is not divergent, Equation (18) should describe the conservation of angular momentum of the material element forming an elastomer. This means that Equation (17) breaks the hypothesis that the local rigid body rotation does not contribute to stress in elastic theory. Under the assumption of classical continuum mechanics in which the element is treated as particle that only has translation, the shear wave defined by Equation (17) shouldn't exist.

**Traditionally defined shear wave**

The shear stress reciprocity is obtained from the assumption that the element forming continuum can be considered as a particle whose rotation can be neglected and its motion can be described by Newton's law. The assumption has been widely accepted by researchers [1,2,5-10]. However, according to the traditional definition of shear wave, the acceleration field and stress field are rotational. That is, the traditional definition of shear wave breaks the assumption of the motion of elements in the classical continuum mechanics. With an open attitude, below I interpret the deformation of elastomer by assuming that the traditional definition of shear wave is objective.

The traditional definition of shear wave considers that the displacement field caused by shear wave is perpendicular to its propagation direction. Suppose a plane shear wave propagates along the $x_1$ coordinate and the displacement is along $x_2$, it is obtained that the strain tensor and stress tensor at one point have only one component:

$$e_{12} = \frac{\partial u_2}{\partial x_1}, \sigma_{12} = \mu \frac{\partial u_2}{\partial x_1}. \tag{19}$$

By decomposing the stress and strain tensors for the traditionally defined shear wave, it can be easily obtained that the traditionally defined shear wave is a superposition of shear deformation and local rigid body rotation. Therefore, the traditionally defined shear wave in elastomer cannot be explained with Equation (10) or Equation (18) due to Equation (10) only considered the shear deformation and Equation (18) only considered the local rigid body rotation. The traditionally defined shear wave should be a superposition of the shear wave described in Equations (10) and (18). As mentioned above, the displacement caused by shear deformation cannot be independently shown in the displacement equation of motion. That is, the shear deformation must company with the volume deformation or local rigid body rotation. it makes no sense to decompose the displacement gradient into deformation and local rigid body rotation, which brings unnecessary confusion. From Equation (17), it is obtained that the shear deformation must company with the local rigid rotation if we believe that an elastic wave can travel in an elastic medium at a shear wave velocity.

In order to balance the moment of momentum caused by the local rigid body rotation, the shear stress reciprocity must be broken, and it should be admitted that the antisymmetric strain originates from local rigid body rotation. It is important to point out that the local rigid-body rotation here should not include the integral rigid-body rotation of the continuum. In the following statement, the continuum is assumed not to rotate as a whole. At this time, the constitutive relation and the elastic tensor which describes resistance to deformation of a material need to be modified appropriately. The study believes that the stress-strain relationship of elastomer should be expressed as follows:

$$\sigma = C : \nabla u . \tag{20}$$

That is, the deformation and local rigid body rotation in the theory of elasticity both contribute stress. Here the local rigid body rotation is called as the torsional deformation. The deformation mentioned below will not be limited to the traditional definition of deformation. For isotropic elastomers, the elastic tensor should be as follows

$$C_{ijkl} = \lambda \delta_{ij}\delta_{kl} + 2\mu \delta_{ik}\delta_{jl} . \tag{21}$$

It is seen that when the displacement gradient has rotation, which corresponds to the material element rotates, the stress tensor is asymmetric. When the displacement field is irrotational, that is, the material elements do not rotate, the stress tensor is symmetric. This means that Equations (20) and (21) contain the case in the theory of elasticity. In fact, the traditional representation of the elastic tensor is also redundant as that the local rigid body rotation does not contribute to the stress has been declared in the theory of elasticity.

As the definition of shear wave breaks shear stress reciprocity, Equation (1) will not be able to describe the motion of the material element corresponding to the shear wave. As the shear wave is related to the rotation of material element, the equation of shear wave should be derived from the angular momentum conservation of the material element. The conservation equation of angular momentum of an elastomer should be expressed as:

$$\int -R \times \frac{1}{2}(\varepsilon : \sigma) \mathrm{d}l = \int R \times \rho \frac{\partial^2 u}{\partial t^2} \mathrm{d}S , \tag{22}$$

where, $R$ is the radius vector of the origin at the centroid, $l$ is the boundary line vector, $S$ is the surface element vector. With the Stokes' theorem, the conservation of angular momentum of material element can be expressed in differential form as follows:

$$-\frac{1}{2}\nabla \times (\varepsilon : \sigma) = \rho \frac{\partial^2 u}{\partial t^2} . \tag{23}$$

For a linear elastic body, the deformation satisfies the superposition principle, and the local rigid body rotation can be regarded as the superposition of two asymmetric deformations. Therefore, the stress tensor $\sigma$ in Equation (23), which is an antisymmetric second tensor, can be expanded to an asymmetric second order tensor, which is determined with Equation (20). The purpose of this paper is to point out that the definition of shear wave is in conflict with the description of material element motion. The relevant issues are not further discussed here.

**Conclusion**

In summary, this study points out that there are mistakes during the derivation of wave equation and the traditional definition of shear wave breaks the shear stress reciprocity. Under the assumption that the motion of the material element forming continuum can be regarded as the motion of a particle, the shear wave should be defined as a wave propagating at shear wave velocity, in which the displacement field caused by shear deformation is neither rotational nor divergent. The definition is different from the traditionally defined shear wave. To some extent, this study challenges the particle description of the motion of material element in classical continuum mechanics. Future study will systematically discuss the problems existing in continuum mechanics.

**Acknowledgments**

The author would like to thank the members of the wave and rock physics research group in Harbin institute of technology and W N Zou from Nanchang University for contributions. This research did not receive any specific grant from funding agencies in the public, commercial, or not-for profit sectors